\documentclass[prl,twocolumn,groupedaddress,superscriptaddress]{revtex4}

\usepackage{graphicx}
\usepackage{color}

\begin{document}

\title{Itinerant magnetic excitations in antiferromagnetic CaFe$_2$As$_2$}

\author{S. O. Diallo}
\affiliation{Ames Laboratory USDOE, Ames, IA 50011 USA}
\author{V. P. Antropov}
\affiliation{Ames Laboratory USDOE, Ames, IA 50011 USA}
\author{T. G. Perring}
\affiliation{ISIS Facility, Rutherford Appleton Laboratory, Chilton, Didcot, Oxon OX11 OQX, United Kingdom}
\affiliation{Department of Physics, University College London, Gower Street, London, WC1E 6BT, United Kingdom}
\author{C. Broholm}
\affiliation{Department of Physics and Astronomy, Johns Hopkins University, Baltimore, MD 21218 USA}
\author{J. J. Pulikkotil}
\affiliation{Ames Laboratory USDOE, Ames, IA 50011 USA}
\author{N. Ni}
\affiliation{Ames Laboratory USDOE, Ames, IA 50011 USA}
\affiliation{Department of Physics and Astronomy, Iowa State University, Ames, IA 50011 USA}
\author{S. L. Bud'ko}
\affiliation{Ames Laboratory USDOE, Ames, IA 50011 USA}
\affiliation{Department of Physics and Astronomy, Iowa State University, Ames, IA 50011 USA}
\author{P. C. Canfield}
\affiliation{Ames Laboratory USDOE, Ames, IA 50011 USA}
\affiliation{Department of Physics and Astronomy, Iowa State University, Ames, IA 50011 USA}
\author{A. Kreyssig}
\affiliation{Ames Laboratory USDOE, Ames, IA 50011 USA}
\author{A. I. Goldman}
\affiliation{Ames Laboratory USDOE, Ames, IA 50011 USA}
\affiliation{Department of Physics and Astronomy, Iowa State University, Ames, IA 50011 USA}
\author{R. J. McQueeney}
\affiliation{Ames Laboratory USDOE, Ames, IA 50011 USA}
\affiliation{Department of Physics and Astronomy, Iowa State University, Ames, IA 50011 USA}

\date{\today}

\begin{abstract}
Neutron scattering measurements of the magnetic excitations in single crystals of antiferromagnetic CaFe$_2$As$_2$ reveal steeply dispersive and well-defined spin waves up to an energy of $\sim$ 100 meV. Magnetic excitations above 100 meV and up to the maximum energy of 200 meV are however broader in energy and momentum than the experimental resolution. While the low energy modes can be fit to a Heisenberg model, the total spectrum cannot be described as arising from excitations of a local moment system.  Ab-initio calculations of the dynamic magnetic susceptibility suggest that the high energy behavior is dominated by the damping of spin waves by particle-hole excitations.
\end{abstract}

\maketitle
%Introduction

 There are indications \cite{Mazin:08,Qiu:08,Matan:08} that superconductivity (SC) in the iron arsenides (Fe-As) family may be driven by a magnetic pairing mechanism, the nature of which remains poorly understood. Recent inelastic neutron scattering experiments have uncovered a magnetic resonance feature in superconducting Ba$_{0.6}$K$_{0.4}$Fe$_2$As$_2$ \cite{Christianson:08}, Ba(Fe$_{0.92}$Co$_{0.08}$)$_2$As$_2$ \cite{Lumsden:08} and Ba(Fe$_{0.95}$Ni$_{0.05}$)$_2$As$_2$ \cite{Chi:08}, that indeed suggests a close relationship between magnetism and superconductivity. The interpretation of these data as excitonic \cite{Eremin:08} or spin wave-like \cite{Chubukov:08} depends crucially on the nature of the Fe-As magnetism, in particular the dimensionality and degree of itinerancy, which we here explore through magnetic neutron scattering from the parent compound CaFe$_2$As$_2$.

Recent experiments probing long wavelength spin waves in antiferromagnetic Fe-As single crystals revealed very large spin wave velocities and anisotropic three-dimensional magnetism \cite{Zhao:08,McQueeney:08}. Another measurement on polycrystalline BaFe$_2$As$_2$ showed excitations as high as 100 meV \cite{Ewings:08}, but the powder averaging inevitably restricts the information that can be extracted from the data.  In this Letter, we report neutron scattering measurements from single crystals of CaFe$_2$As$_2$ probing magnetic excitations up to 200 meV. Steeply dispersive spin waves, consistent with previous measurements, were observed up to $\sim$ 100 meV. Beyond 100 meV, and up to the maximum observed magnetic excitation of $\sim$ 200 meV, damping is substantial and collective excitations are not clearly visible in the data.  Linear response calculations of the dynamic susceptibility reveal that the high energy antiferromagnons are strongly affected by particle-hole excitations (Landau damping), indicating that the spin excitations have an itinerant nature over a large fraction of the Brillouin zone. A Heisenberg model can account for magnetic excitations below 100 meV and was used to extract effective low energy exchange parameters and an energy dependent spin wave relaxation rate. The local moment (Heisenberg) description of magnetism in this class of materials is however, limited to small wave-vectors and low energies with substantial effects of electron-itinerancy beyond 100 meV.    
%Experiment

CaFe$_2$As$_2$ is a non-superconducting parent compound of the new family of iron arsenide based superconducting materials \cite{Ni:08}. CaFe$_2$As$_2$ undergoes a paramagnetic to antiferromagnetic (AFM) transition simultaneously with a tetragonal (I4/mmm) to orthorhombic (Fmmm) structural transition below $T_N=T_S=$ 172 K \cite{Ni:08}. The orthorhombic lattice parameters are $a=$ 5.51 {\AA}, $b=$ 5.45 {\AA}, and $c=$ 11.67 {\AA} at $T=10$ K. The magnetic structure is a collinear antiferromagnet with propagation vector ${\bf Q}_{AFM}=(1 0 1)$. Throughout this paper, wavevectors are specified in the orthorhombic reciprocal lattice. The Fe ordered moment is 0.80(5)$\mu_B/$Fe  directed along the orthorhombic $a$ axis \cite{Goldman:08_1}.

  The measurements were performed on the MAPS time-of-flight chopper spectrometer at the ISIS neutron facility. The sample consisted of approximately 400 crystals grown from tin flux, with a total mass $\sim$ 2 grams. These were co-aligned with an angular mosaic width of 1.5 degrees and a $[H0L]$ scattering plane. Crystals were etched in concentrated hydrochloric acid to remove Sn flux from the surface. Magnetic Brillouin zone (BZ) centers are located at ${\bf Q}_{AFM}=(H=2m+1,K=2n,L=2j+1)$, $m,n,j$ being integers. The sample was mounted on the cold finger of a closed-cycle refrigerator and cooled to $T=$ 10.0(5) K for measurements at four incident energies; $E_i=$ 100, 180, 300, and 450 meV with the incident beam along [100] and [001].

The quantity of interest is the dynamical structure factor $S({\bf Q},\omega)$, which is related to the double differential cross-section via, $S({\bf Q},\omega)=({k_i}/{k_f}) d^2\sigma/d\Omega_f dE_f$. Here subscripts $i$ and $f$ label the initial and final neutron wave vector $k$, $\Omega$ is the solid angle of scattering, and $E_f$ is the scattered neutron energy. Incoherent nuclear scattering from a standard vanadium sample was used to normalize the data so that $S({\bf Q},\omega)$ is reported in units of mbarn sr$^{-1}$ meV$^{-1}$ per formula unit of CaFe$_2$As$_2$ throughout. We used the MSLICE program \cite{Coldea:04} to visualize the data and to take one and two dimensional cuts through main crystallographic symmetry directions for subsequent data analysis with TOBYFIT \cite{Tobyfit:04}. Symmetry equivalent cuts and slices were added to improve statistics. We used TOBYFIT to simulate the scattering cross-section for damped antiferromagnetic Heisenberg spin waves convoluted with the instrumental resolution function for comparison with the experiment. We note that orthorhombic twinning is present and accounted for in the simulations.

\begin{figure}
\begin{tabular}{c c}
\includegraphics[width=.5\linewidth]{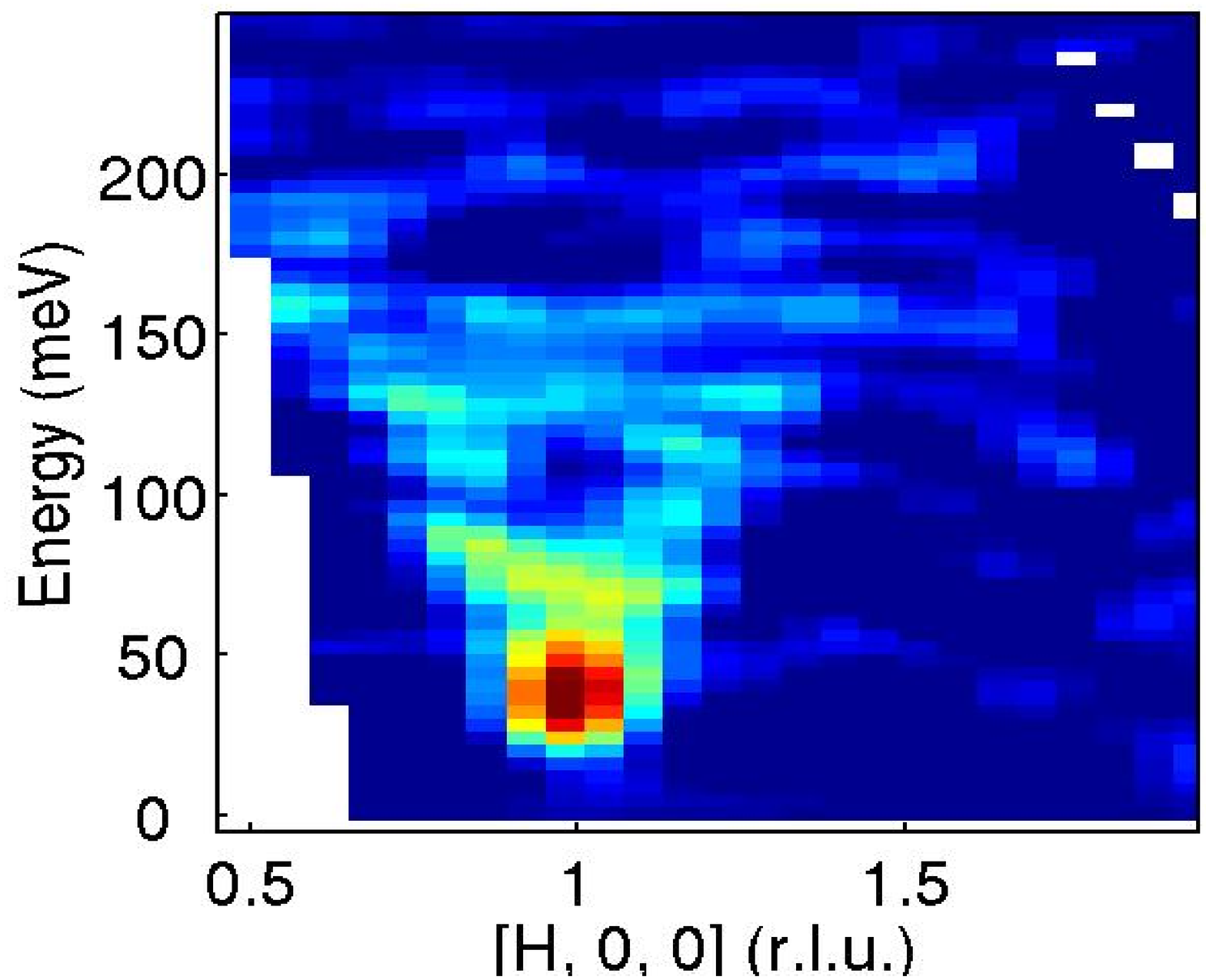}&\hspace{-0.25cm}\includegraphics[width=.5\linewidth]{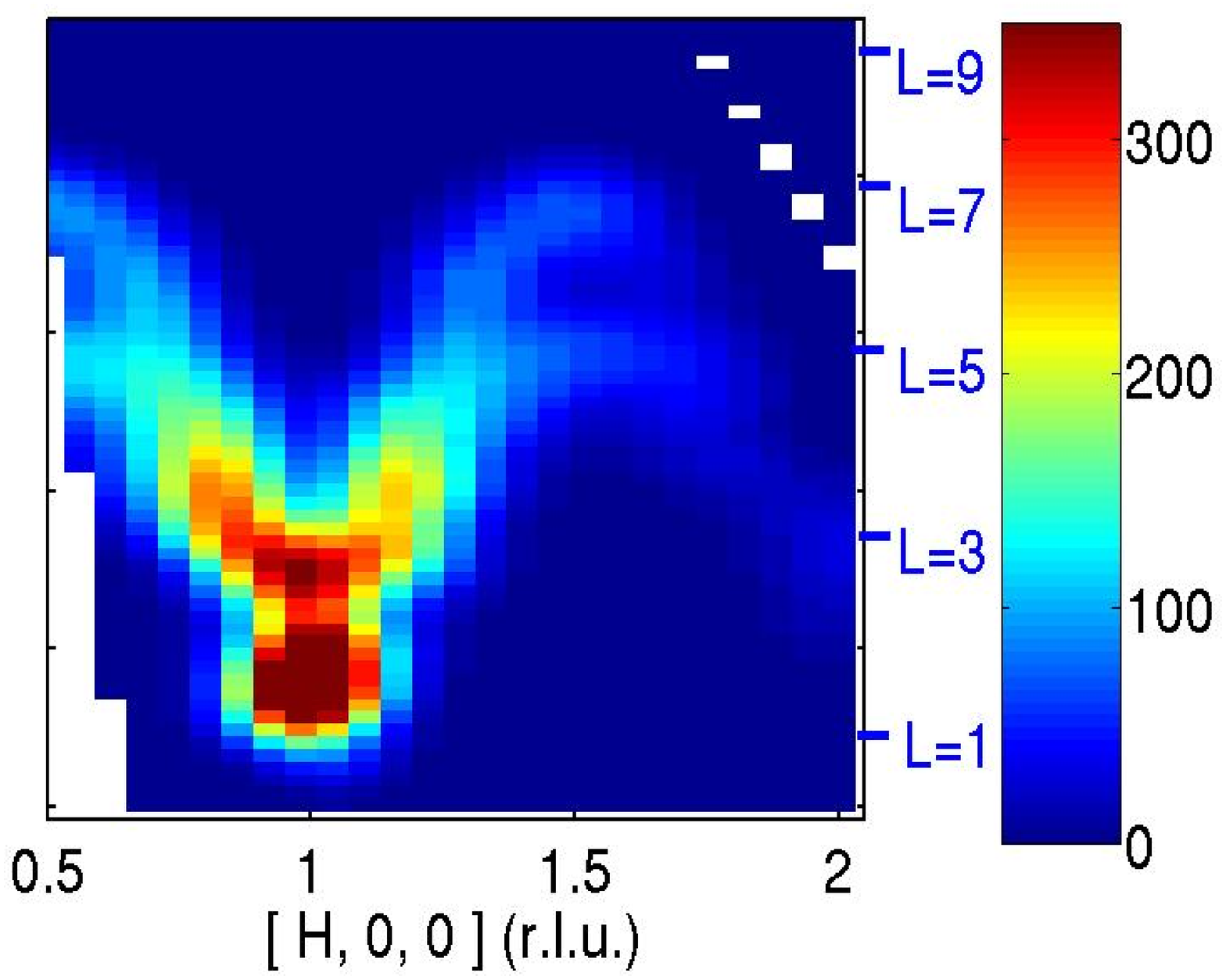}\end{tabular}
\caption{\footnotesize Observed (left) and calculated (right) magnetic excitations in CaFe$_2$As$_2$ single crystals at temperature $T=$ 10 K with an incident energy $E_i=450$ meV. The projections are in the scattering plane formed by the energy transfer axis and $(H,0,0)$ direction and averaged over $-0.1<K<0.1$. An estimated background away from the magnetic zone center averaged over $1.8<H<2.0$ and $-0.25<K<0.25$ has been subtracted from the data. The intensities have been multiplied by the energy transfer for viewing purposes.  Due to the fixed crystal orientation with incident beam along $L$, the $L$ component of the wavevector varies with energy transfer. The calculation is from a Heisenberg spin wave model with a small damping parameter ($\Gamma$=3 meV) (see text).} 
\label{fig_dispr}
\end{figure}

Fig. \ref{fig_dispr} shows a projection of the $E_i$ = 450 meV data along the $H$-direction and near the $(10L)$ antiferromagnetic Bragg peaks as compared to the structure factor calculated from a Heisenberg spin wave model with small damping (within instrumental precision) using experimentally determined parameters (discussed below). The sizeable exchange interaction between planes leads to variations in the structure factor along $L$. The data indicate magnetic excitations beyond 150 meV and extending perhaps as high as 200 meV. Excellent agreement between model and data was achieved though with considerable spin wave damping for energies beyond 100 meV as will be discussed later.

\begin{figure}
\begin{tabular}{c c}
\includegraphics[width=.5\linewidth]{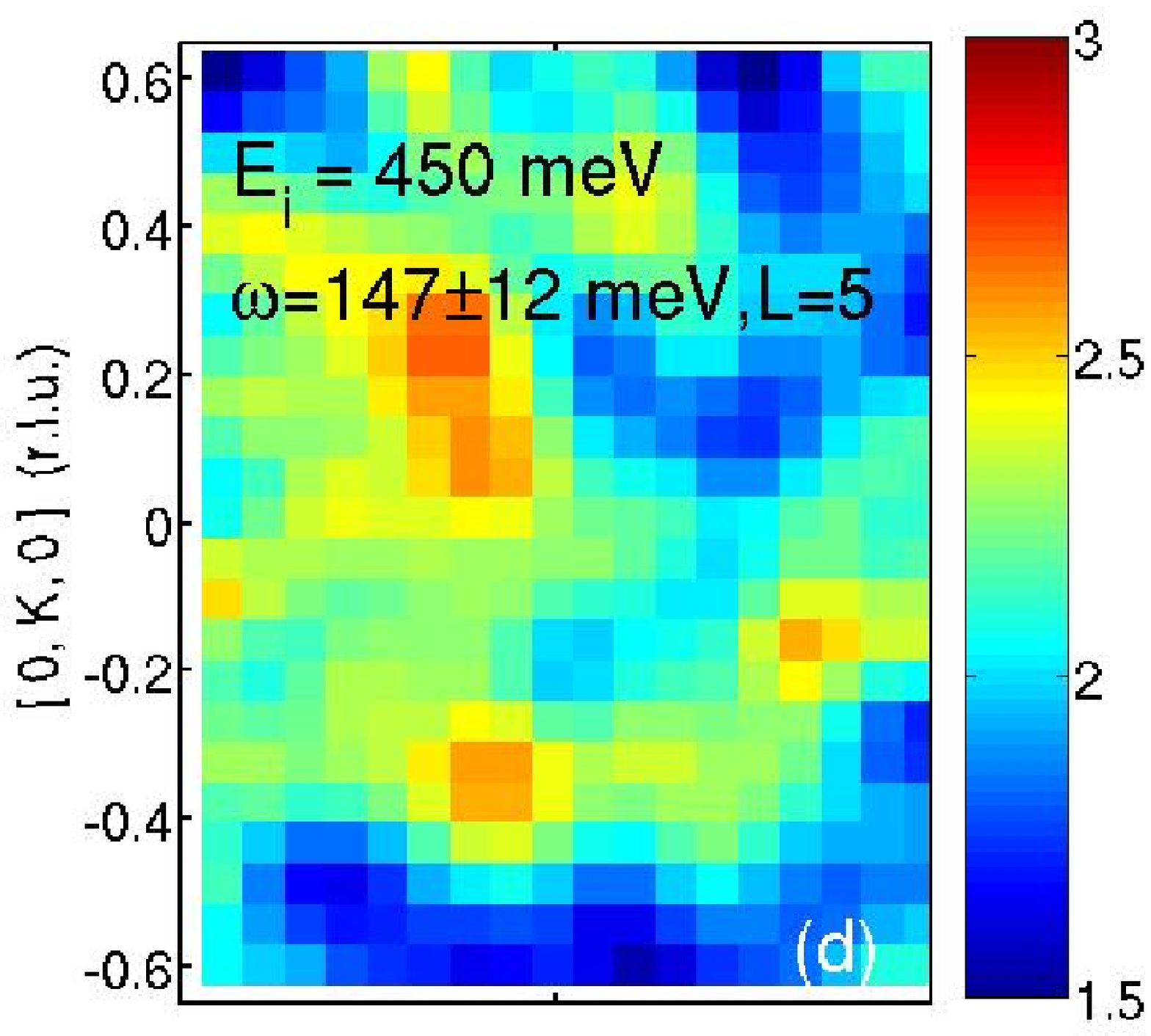}&\includegraphics[width=.5\linewidth]{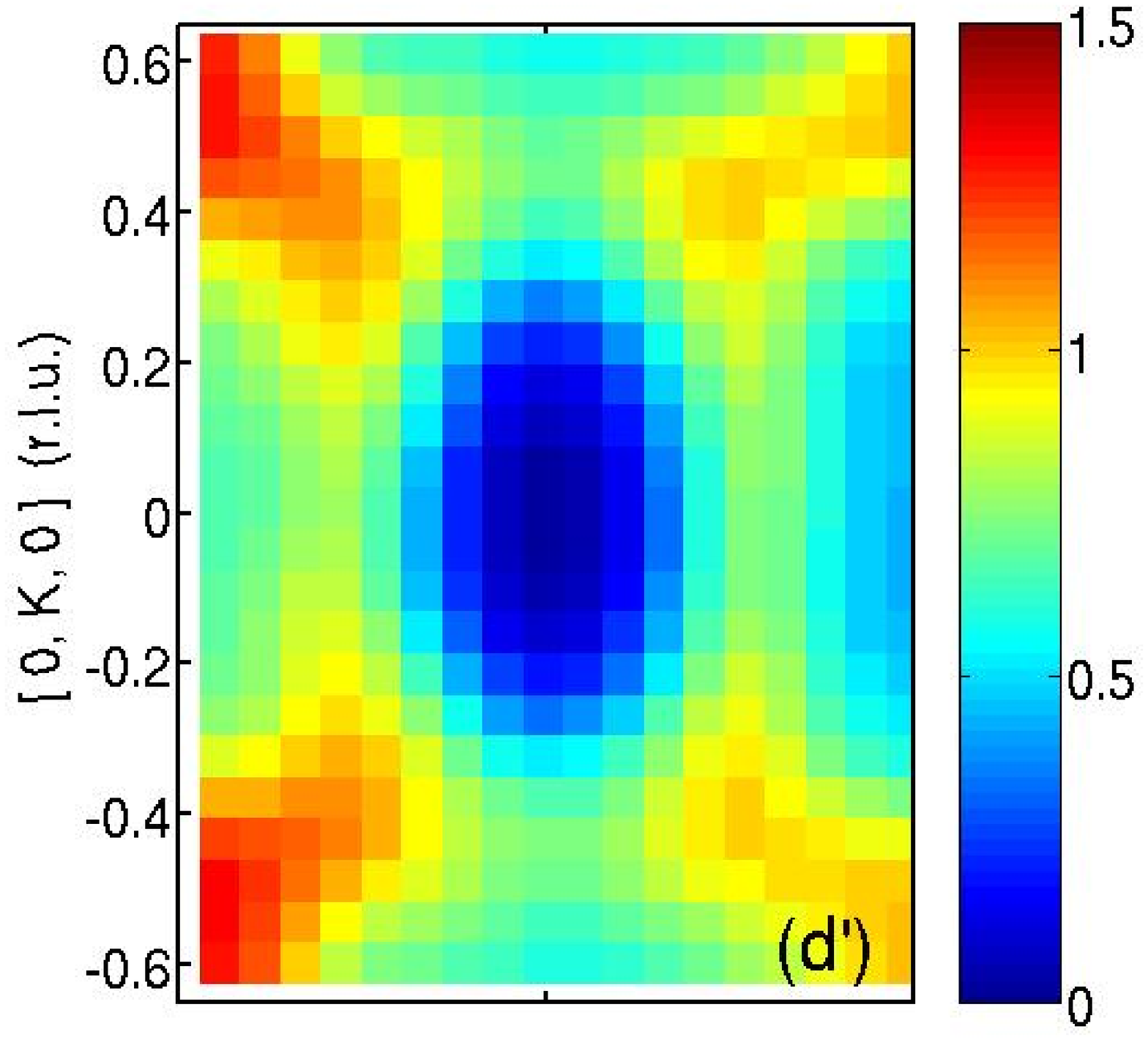} \\
\includegraphics[width=.5\linewidth]{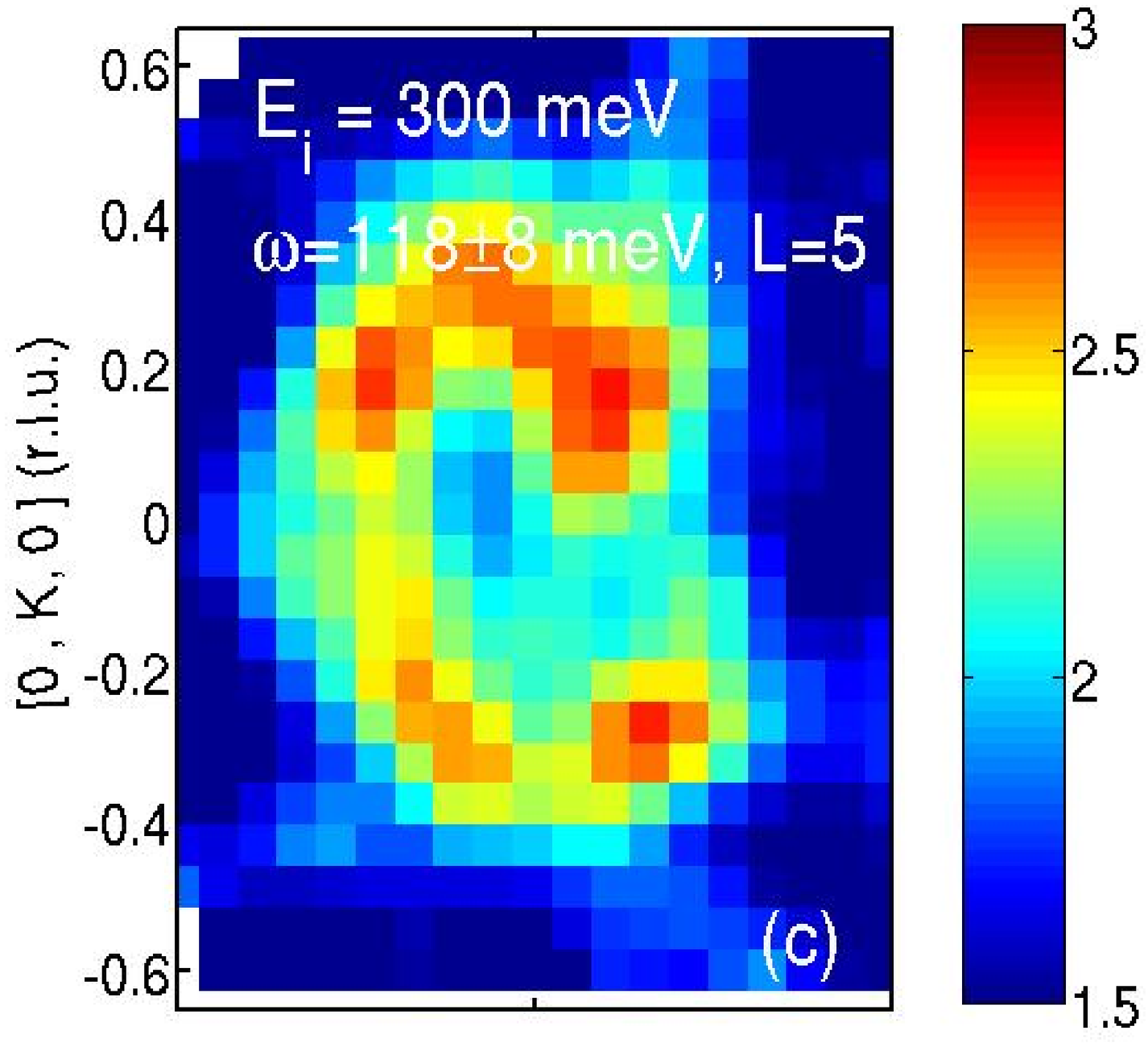}&\includegraphics[width=.5\linewidth]{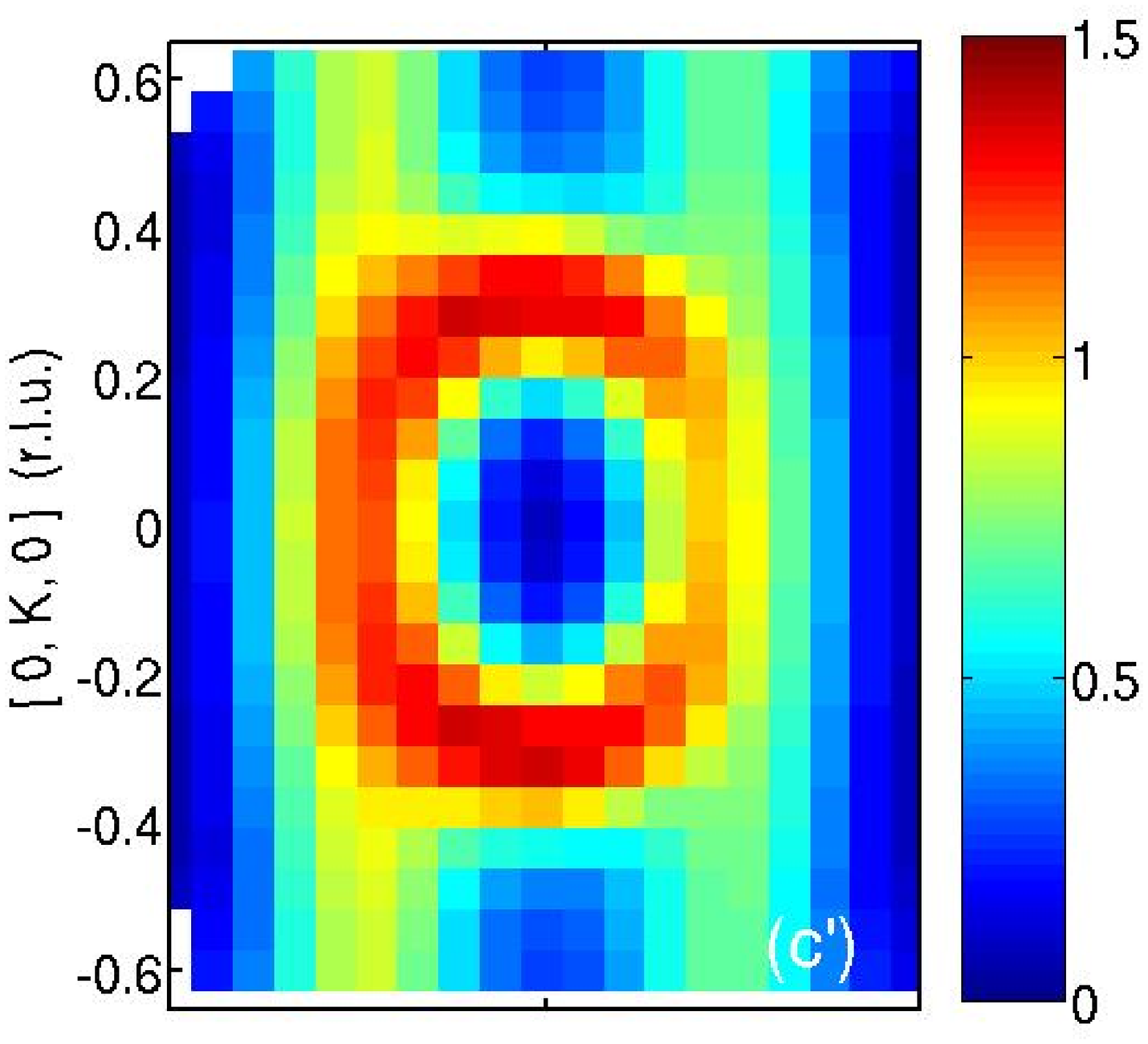}\\
\includegraphics[width=.5\linewidth]{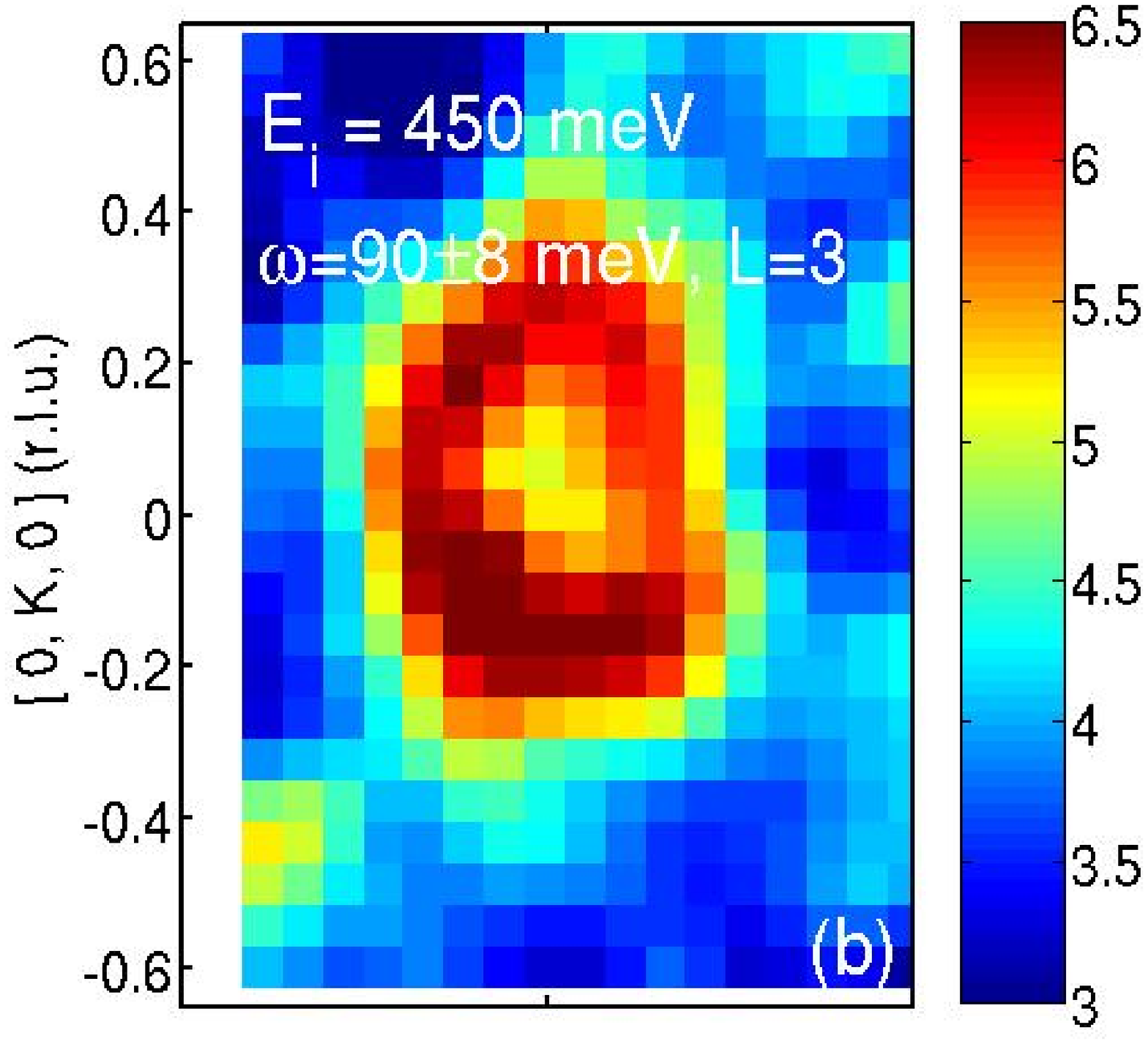}&\includegraphics[width=.5\linewidth]{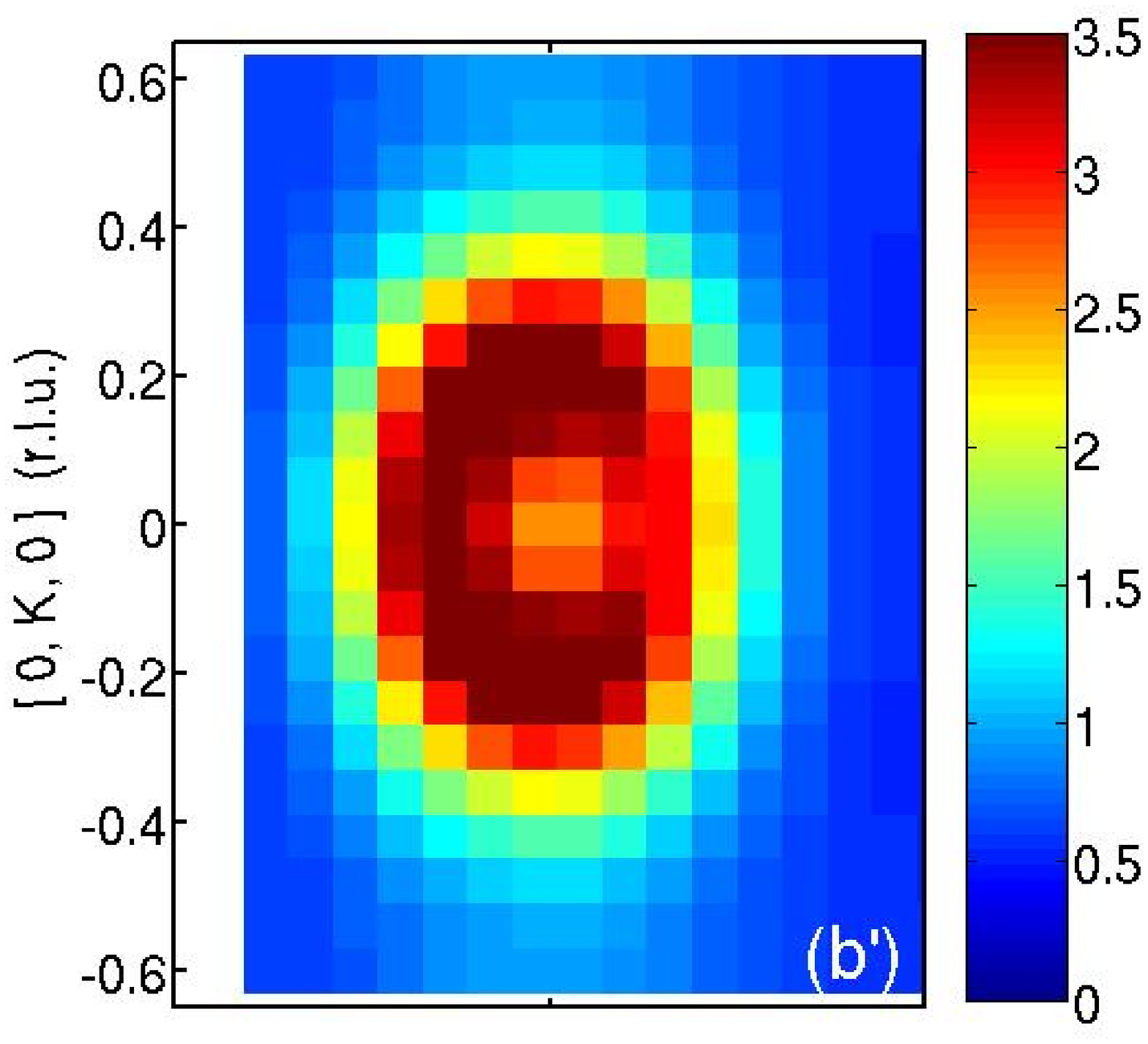}\\
\includegraphics[width=.5\linewidth]{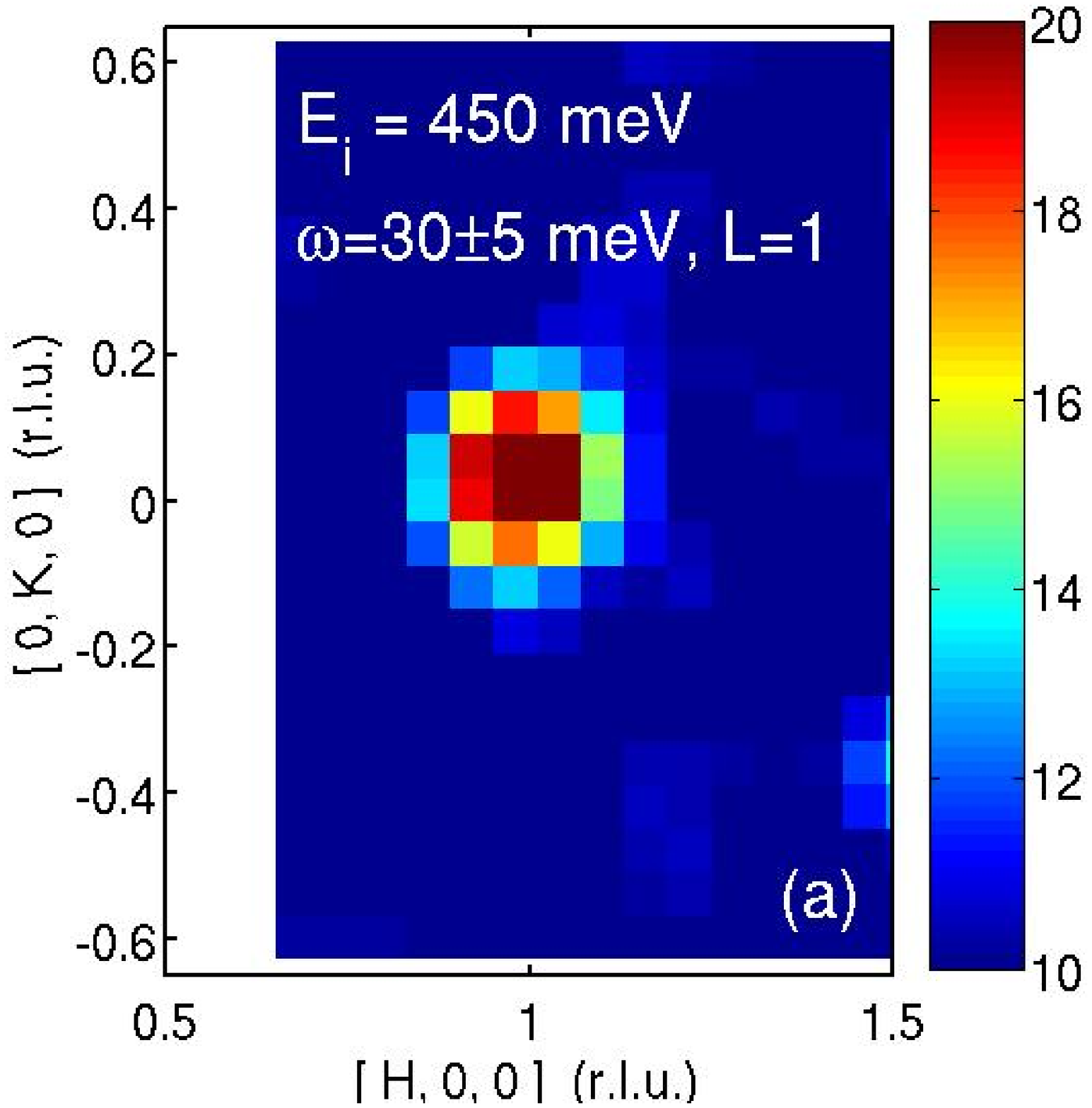}&\includegraphics[width=.5\linewidth]{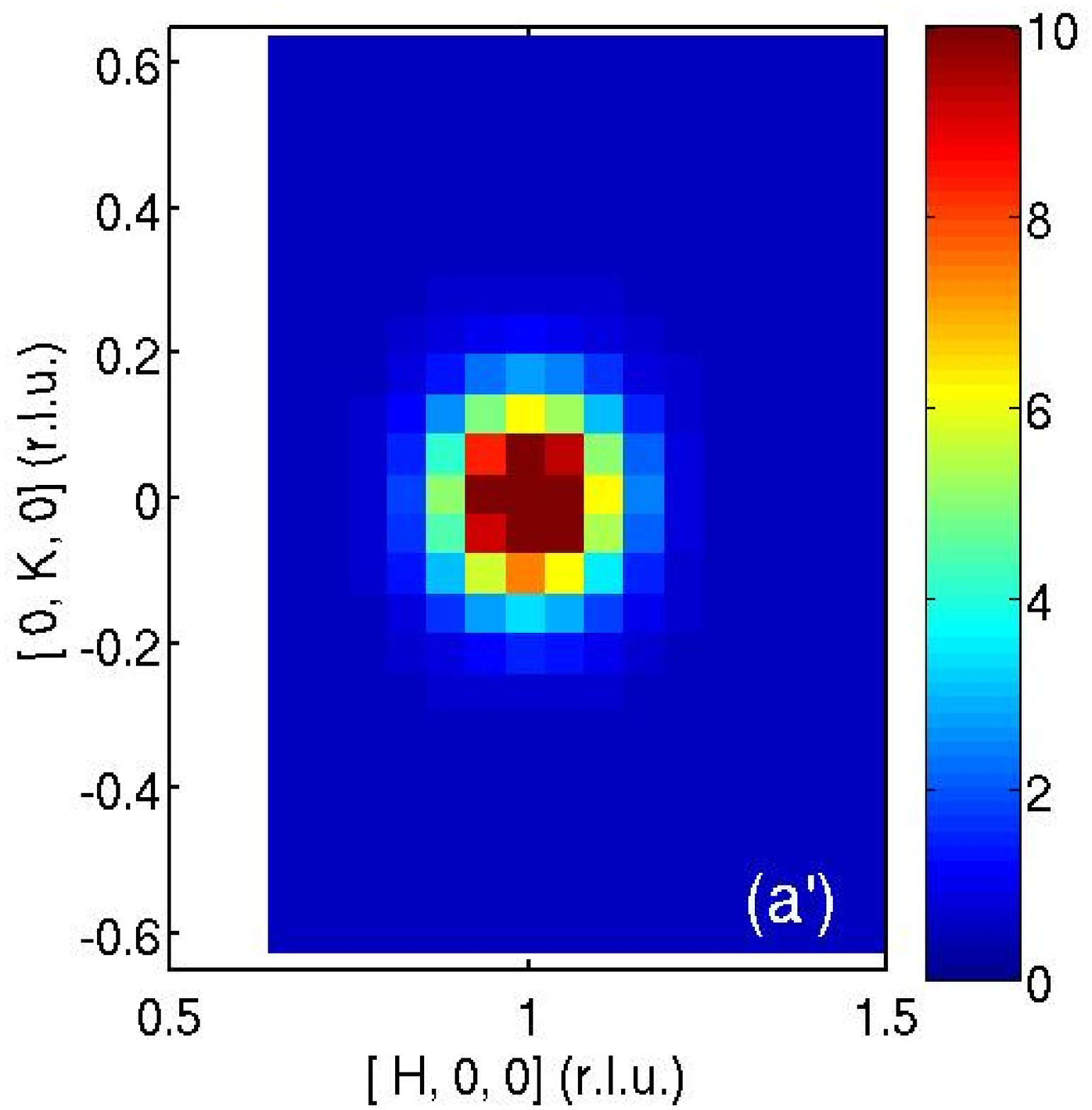}
\end{tabular}
\caption{\footnotesize Constant energy slices through spin waves in CaFe$_2$As$_2$ at $T=$ 10 K as observed on MAPS (left) (intensity shown in absolute units from background level) compared with calculations using the fitted exchange parameters in the text (right) with small damping ($\Gamma$=3 meV) (see text).} 
\label{fig_rings}
\end{figure}

\begin{figure}
\includegraphics[width=1\linewidth]{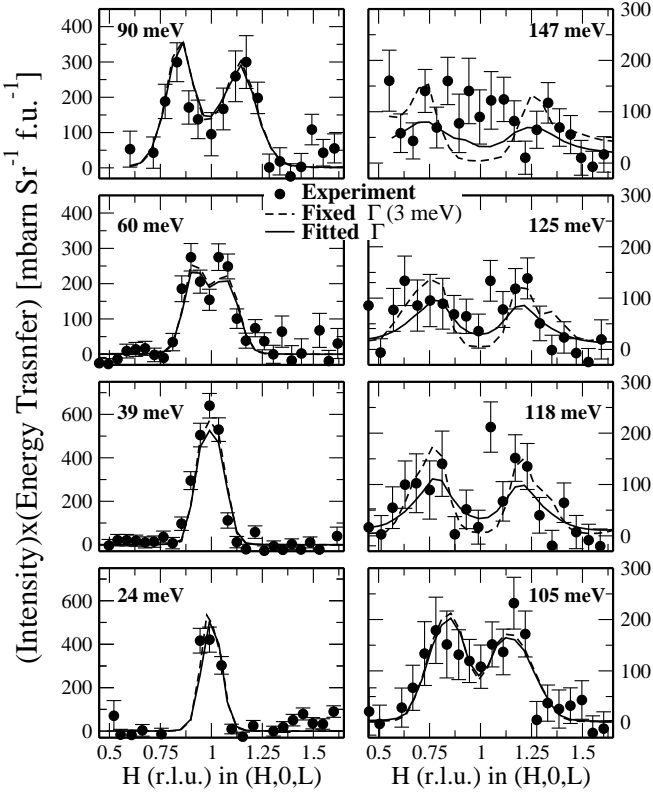}
\caption{\footnotesize Intensity of observed spin wave excitations in CaFe$_2$As$_2$ along $[H00]$ direction at $T=$ 10 K, compared with the resolution convoluted spin-wave calculations including damping shown in Fig. 5.  The calculations with energy independent damping ($\Gamma$ held fixed at 3 meV) are also shown for comparison.}
\label{fig_cuts}
\end{figure}

Fig. \ref{fig_rings} shows constant energy slices taken in the $HK$-plane at specific energy transfers such that $L=2j+1$ [i.e. centered at ${\bf Q}_{AFM}=(1,0,L=2j+1)$]. The data clearly show rings of scattering as expected for constant energy slices through a conical dispersion surface emerging from ${\bf Q}_{AFM}$.  Ring-like features are well-resolved up to $\sim$ 100 meV [see Fig. \ref{fig_rings}(c)]. Beyond 100 meV [Fig. \ref{fig_rings}(d)], the data show smeared out and broadened features centered at ${\bf Q}_{AFM}$. Fig. \ref{fig_cuts} emphasizes this observation by showing several cuts along $H$ at energies corresponding to $L=2j+1$. Well separated spin-wave peaks are observed between 60 and 118 meV. Above 100 meV, splitting is however not clearly observed due to weakening and/or broadening of the scattering. 

\begin{figure}
\includegraphics[width=1\linewidth]{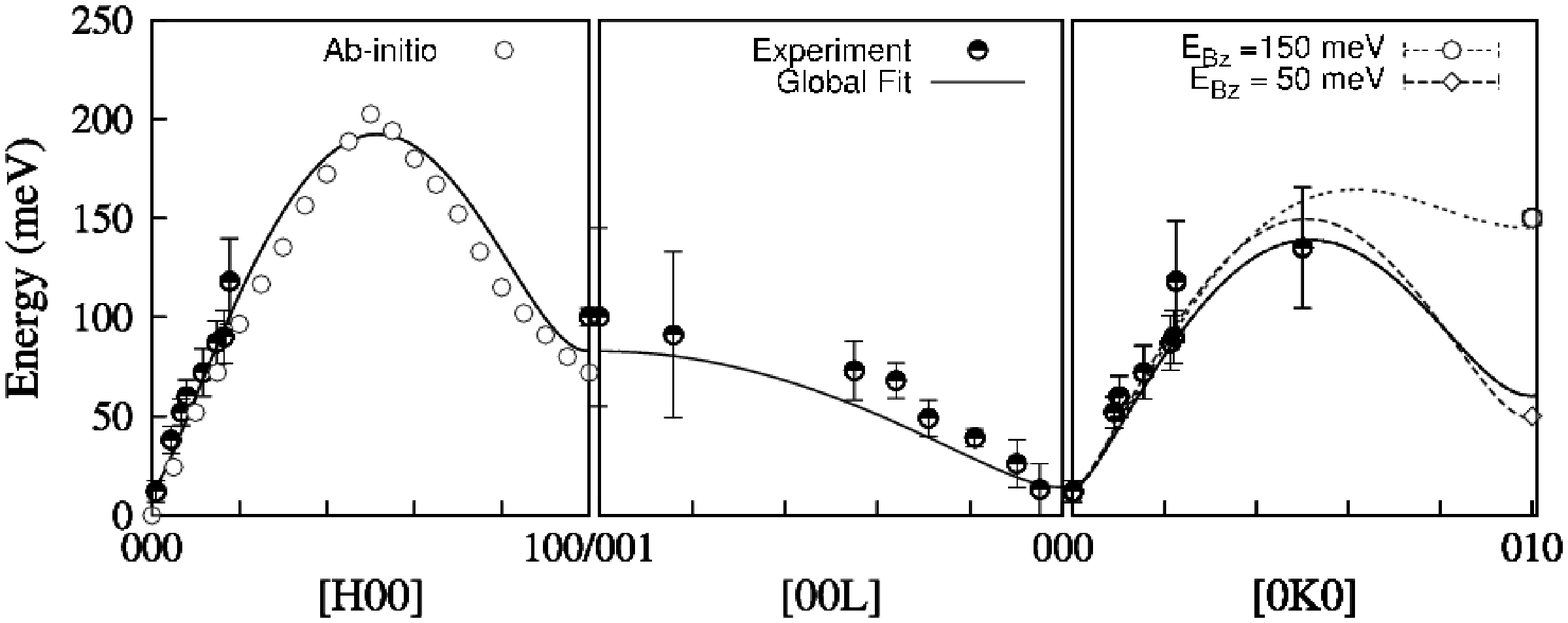}
\caption{\footnotesize The dispersion of magnetic excitations in CaFe$_2$As$_2$: theory (lines) and experiment (solid circles). The dashed black lines are from Eq. \ref{eq_model} using exchange parameters derived from band structure calculations (see Ref. [\onlinecite{McQueeney:08}]). The solid red lines represent the best global fit of Eq. \ref{eq_model} to the data. Variations of the fits with changing BZ boundary energy are shown for $E_{BZ}$=50 and 150 meV. Also shown with open circles are the location of maxima in Im$\{\chi\left({\mathbf q},\omega\right)\}$ from ab-initio band structure results.}
\label{fig_fits}
\end{figure}

The data cuts, such as those in Fig. \ref{fig_cuts}, were fit to Gaussian lineshapes in order to determine the relative position of the peaks with respect to the BZ center.  The peak positions were used to construct the spin wave dispersion relation along major symmetry directions (Fig. \ref{fig_fits}.) These data are compared to the dispersion relation in the linear approximation for a Heisenberg model with nearest-neighbor (NN) and next-nearest-neighbor (NNN) interactions. The associated dispersion relations are given by
\begin{equation}
E({\bf q}) = \sqrt{A_{\bf q}^2-B_{\bf q}^2},
\label{eq_model}
\end{equation}
\noindent where
\begin{eqnarray}
A_{\bf q} &=& 2\langle S\rangle [J_{1b}(\cos(\pi k)-1)+J_{1a}+J_{1c}+2J_2+D]\nonumber\\
B_{\bf q} &= & 2\langle S\rangle [J_{1a}\cos(\pi h)+2J_2\cos(\pi h) \cos(\pi k)\\
         & &+J_{1c} \cos(\pi l)]\nonumber
\label{eq_model_I}.
\end{eqnarray}
\noindent Here the NN Heisenberg exchange constants are denoted $J_{1a}$, $J_{1b}$, and $J_{1c}$ and the in-plane NNN exchange constant $J_{2}$.    $\langle S\rangle$ is the ordered spin,  ${\bf q}=(hkl)={\bf Q}-{\bf Q}_{AFM}$ is the reduced wave vector, with ${\bf Q}$ the wave vector transfer to the sample, and $D$ is a single ion uniaxial anisotropy constant. The corresponding equations for the spectral weight of the spin waves used in the simulation together with Eq. \ref{eq_model} are given explicitly in \cite{Ewings:08}.
  
Heavy damping at high energies renders the Brillouin zone boundary spin wave energies ill defined. As a result, only three independent energy scales can be extracted from the data and these are $\bar{J_1}=(J_{1a}+J_{1b})/2$, $\tilde{J_2}=J_2+\delta J_1/2$, and $J_{1c}$. Here $\delta J_1=(J_{1a}-J_{1b})/2$, is related to the putative zone boundary energy, which when $J_{1c}<<J_{1a}$ and $J_{1c}<<J_2$ is given by $E({\bf q}=(010))=4\langle S\rangle \sqrt{2\delta J_1(2\tilde{J}_2-\bar{J_1})}$. Fits with full resolution convolution to the cuts in Figs. 1-4 yield $\langle S\rangle \bar{J_1}=21.9(9)$~meV, $\langle S\rangle \tilde{J_2}=34.8(4)$~meV, and $\langle S\rangle J_{1c}=4.5(1)$~meV. We maintained the zone center gap at $\Delta=7$~meV in accordance with our higher resolution low energy data\cite{McQueeney:08} and this corresponds to  $\langle S\rangle D\approx (\Delta^2/8\langle S\rangle(\bar{J}_{1}+2\tilde{J_2}+J_{1c}))=0.063$~meV. We obtain, $g\langle S\rangle\mu_{B} = 0.8(1) \mu_{B}$ from the intensity and this is consistent with the observed ordered moment.  If 50 meV$<E({\bf q}=(010))<150$~meV these results indicate that 2 meV$<\langle S\rangle \delta J_1<15$~meV such that 24 meV$<\langle S\rangle J_{1a}<37$~meV, 7 meV$<\langle S\rangle J_{1b}<20$~meV, and 28 meV$<\langle S\rangle J_2<34$~meV. The fitted values are consistent with the requirement that $\langle S\rangle (J_{1a}+J_{1b}) =2\langle S\rangle \bar{J}_1=44$~meV$< 4\langle S\rangle J_{2}=124$~meV$\pm 12$~meV for the \lq stripe' AFM structure.  The results also confirm that $\langle S \rangle J_{c}=4.5(1)$~meV is substantial and the magnetic system is three-dimensional (our previous measurements \cite{McQueeney:08} could only establish a lower bound for $\langle S \rangle J_{c}$).  Fig. \ref{fig_fits} summarizes the fit results.

The fits described above do not require substantial damping at low energies and the calculated intensities agree reasonably well there with no adjustable scale factor (see Fig. \ref{fig_cuts}). This indicates that the full static moment participates in long wavelength collective (spin wave) excitations. To account for the data beyond 100 meV it is however, necessary to introduce considerable spin wave damping. For simplicity the spin wave relaxation rate, $\Gamma(E)$, was assumed to depend only on energy and was treated as piecewise constant in 10 energy bands. The simulations shown in Fig.~3 include energy dependent damping required to fit the data and Fig.~5 shows the corresponding $\Gamma (E)$. This analysis provides evidence for Landau damping in the long wavelength limit and strong intrinsic broadening beyond 100 meV. Both findings are consistent with expectations for an itinerant antiferromagnet. The situation is remarkably similar to underdoped  YBa$_2$Cu$_3$O$_{6.35}$ ($T_c=$ 18 K) where there is an onset of damping and reduced spectral weight as compared to conventional spin wave theory beyond the 150-200 meV pseudo-gap \cite{Frost:07}. 

\begin{figure}
\includegraphics[width=0.8\linewidth]{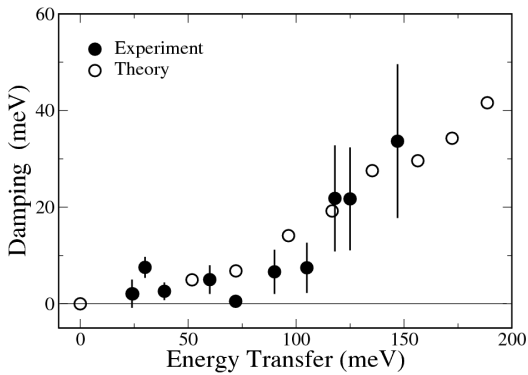}
\caption{\footnotesize Spectral half width at half maximum derived from band-structure calculated peaks in Im$\{\chi\left({\mathbf q},\omega\right)\}$ (open circles) compared to the spin-wave relaxation rate, $\Gamma (E)$ required to fit the neutron scattering data (solid circles).}
\label{fig_wq}
\end{figure}

To better understand the origin of this broadening, theoretical calculations of the dynamic spin susceptibility $\chi \left({\mathbf q}, \omega \right)$  were performed using linear response density functional approach \cite{Zein:01}. The calculated magnetic moment on the Fe atom is 0.95$\mu _{B}$.  Fig. \ref{fig_fits} shows a remarkable agreement between the calculated spin wave spectrum, as derived from the peak position of Im$\{\chi\left({\mathbf q},\omega\right)\}$ along the $H$-direction, the raw experimental peak positions (solid circles), and the fitted dispersion relation. The half-width at half maximum of the calculated peaks, are compared to the spin-wave relaxation rate extracted from the scattering data in Fig. \ref{fig_wq}. Strong line broadening is apparent beyond $\omega \sim$ 100 meV (or reduced $q(100) \sim$ 0.2) and persists up to a maximum spin wave energy of 200 meV where $\Gamma / \omega$ approaches 25\%.  The calculations indicate the presence of incoherent spin fluctuations across most of the BZ consistent with the experimental data.  Numerous band structure calculations indicate that the Fermi level sits in a region of  relatively low density of states (pseudo-gap) approximately 0.1-0.2 eV wide.  The pseudo-gap may serve as a source of stability for local moment collective excitations occurring at small $q < 0.2$ where a Heisenberg model description is justified. At higher energies, ab-initio calculations are best described as gradual evolution of the spin dynamics into the Stoner continuum. 

%Conclusions
The present results favor an itinerant model for magnetic excitations in the parent CaFe$_2$As$_2$ compound, as opposed to a local moment Heisenberg model. At small $q$, an effective NNN Heisenberg model is always an appropriate description for either a local moment or itinerant system. The real applicability of such a model should be tested at larger $q$ ($q>0.2$). At these $q$ both experiment and theory  show the appearance of substantial damping. Thus, the parent compounds already manifest strong coupling of magnetism and charge carriers in the form of an evolution of Landau damping to a full particle-hole continuum. This general  remark should also apply to the doped superconducting compositions where AFM ordering is suppressed but essential AFM correlations persist up to high energies \cite{Christianson:08}. In these compositions, it appears that strong Landau damping of the magnetic excitations occurs at much lower energies ($\sim$ 10 meV) and the SC gap suppresses damping.

In summary, magnetic excitations in antiferromagnetically ordered CaFe$_2$As$_2$ have been observed up to 200 meV. The features of the highly dispersive spin waves are consistent with local moment magnetism for energies below $\sim$ 100 meV and can be represented by a NNN Heisenberg model there. For energies above $\sim$ 100 meV, the excitations are strongly damped indicating presence of particle-hole excitations. CaFe$_2$As$_2$ is thus best described as an itinerant three-dimensional antiferromagnet.

%Acknowledgements

Work at the Ames Laboratory and the Johns Hopkins University was supported by the Department of Energy, Basic Energy Sciences under Contract No. DE-AC02-07CH11358 and DE-FG02-08ER46544 respectively. Technical assistance of ISIS staff is gratefully acknowledged.

\bibliographystyle{apsrev}

\end{document}